\documentclass[conference]{IEEEtran}
\IEEEoverridecommandlockouts
\usepackage{cite}
\usepackage{amsmath,amssymb,amsthm,amsbsy,amsfonts,amsopn,amstext,amsxtra,euscript,amscd,graphicx}
\usepackage{mathrsfs,bm}
\usepackage{algorithmic}
\usepackage{graphicx}
\usepackage{textcomp}
\usepackage{xcolor}
\usepackage{float,boxedminipage}

\newtheorem{definition}{Definition}

\def\BibTeX{{\rm B\kern-.05em{\sc i\kern-.025em b}\kern-.08em
		T\kern-.1667em\lower.7ex\hbox{E}\kern-.125emX}}
\begin{document}

\title{Information-Theoretic Distributed Point Functions with Shorter Keys
    \thanks{Hang Deng (denghang2022@shanghaitech.edu.cn.) is with the
        Institute of Mathematical Sciences, ShanghaiTech University, Shanghai, China.
        Liang Feng Zhang (zhanglf@shanghaitech.edu.cn) is with the School of Information Science and Technology, ShanghaiTech University, Shanghai, China. 
        This work was supported in part by the National Natural Science Foundation of China (No. 62372299)}}

\author{Hang Deng and Liang Feng Zhang}

\maketitle

\begin{abstract}
    A \(t\)-private \(n\)-server Information-Theoretic Distributed Point
    Function (\((t, n)\)-ITDPF) allows one to convert any point
    function \(f_{\alpha, \beta}(x):[N]\rightarrow \mathbb{G}\)
    into \(n\) shares (secret keys), such that each server can compute an additive share of \(f_{\alpha,\beta}(x)\)  with a key while
    any \(\leq t\)    servers learn absolutely no information about the function.
This paper constructs  a novel share conversion   based on the 
 private information retrieval (PIR) of Ghasemi, Kopparty, and Sudan (STOC 2025)
and proposes    a perfectly secure \(1\)-private
     ITDPF  with output group \(\mathbb{G}= \mathbb{Z}_p\), where
      \(p\) can be any prime.   
    Compared with the existing perfectly secure ITDPFs for the same output group,
the proposed ITDPF  is  more efficient with asymptotically shorter secret keys. 
\end{abstract}

\section{Introduction}
\label{sec:Introduction}

A \(t\)-private  \(n\)-server \textit{Distributed Point Function}  (\((t,n)\)-DPF) \cite{GI14,BGIK22}  can convert any point function \(f_{\alpha,\beta}: [N]\rightarrow \mathbb{G}\)
(i.e., a function such that \(f_{\alpha,\beta}(\alpha)=\beta\) and
\(f_{\alpha,\beta}(x)=0\) for all \(x\neq \alpha\))
  into \(n\) shares (called {\em secret keys}) \(k_0,\ldots,k_{n-1}\) such that
every key \(k_i\) enables the computation of
an additive share of \(f_{\alpha,\beta}(x)\) but any \(\leq t\) shares leak no
information about the function.
In particular,  the range \(\mathbb{G}\) of the function
is an Abelian group  and called the {\em output group}.
The communication efficiency of a DPF may be measured by its {\em key size}, i.e.,
the maximum size of the \(n\)  keys  \(k_0,\ldots,k_{n-1}\), as a function of the domain size \(N\).
Ideally,  DPFs with shorter keys are preferred  for any given  
\(t, n\) and     output group \(\mathbb{G}\).

DPFs have numerous applications, e.g.,  in constructing
secure multiparty computation (MPC) protocols and specifically  Private Information Retrieval (PIR) {~\cite{CGKS95}} protocols \cite{Zhang26}, \cite{KZW26}.
While {\em computational} DPFs \cite{GI14} are secure against the collusion of any $t$ polynomial-time servers,  
{\em Information-Theoretic} DPFs (ITDPFs) \cite{BGIK22}
offer stronger security by tolerating  computationally unbounded servers and are our focus in this paper.

 Boyle et al. \cite{BGIK22} and  Li et al. \cite{LKZ25} constructed    ITDPFs
  (see {\sc Table \ref{tab:itdpf}}) that gave various trade-offs among the security, the key size, the number of servers,
and the generality of output groups. In particular, for $\mathbb{G}=\mathbb{Z}_p$, where 
$p$ may be any prime,
 Li  et al.   \cite{LKZ25}   constructed  a {\em perfectly} secure $(t,d(t+1))$-ITDPF with
          {\em polynomial} key size.
 Boyle et al. \cite{BGIK22} and  Li et al.  \cite{LKZ25} 
 proposed  {\em  perfectly} secure $(1,4)$-ITDPF
 and  $(1,2n_r)$-ITDPF with {\em subpolynomial} key sizes
\(2^{O(\nu_2(N))}\) and   $2^{O(\nu_r(N))}$  respectively, where
$O$ may hide a factor in $p$ and 
\begin{align}
\nu_r(N)=(\log N)^{1/r}(\log\log N)^{1-1/r},  
\end{align}

 \begin{align}
    \label{equ:n_value}
    n_r=\left\{
    \begin{array}{ll}
        2,                     & r=1;                  \\
        3^{r/2},               & 1<r\leq 103,2|r;      \\
        8\cdot 3^{(r-3)/2},    & 1<r\leq 103,2\nmid r; \\
        (3/4)^{51}\cdot 2^{r}, & r\geq 104.
    \end{array}
    \right.
\end{align}
  Boyle et al.     \cite{BGIK22} and Li et al. 
\cite{LKZ25} also developed {\em statistically} secure  ITDPFs with subpolynomial key sizes, which 
however trade off the stronger perfect security for a smaller number of required servers.  
 
Underlying \cite{BGIK22,LKZ25} is   a framework that firstly  builds    share conversions  from PIR and then
represents point functions as   bilinear functions  of  share conversions.
In particular, the key size  of the resulting ITDPF  is propositional to the communication complexity of
the underlying PIR. Recently, 
Ghasemi et al. \cite{GKS25} proposed the state-of-the-art 1-private PIR, which 
makes it promising  to construct ITDPFs with asymptotically  shorter secret keys
using the framework of \cite{BGIK22,LKZ25}.

\vspace{-2mm}
 \begin{table}[H]
    \caption{\bf  Perfectly secure $(t,n)$-ITDPFs with  output group $\mathbb{Z}_p$}
    \vspace{-2mm}
    \label{tab:dpf-comparison}
    \centering
    \begin{tabular}{|c|c|c|c|}
        \hline
          Constructions                & $t$  & $n$ &    Key size                    \\
        \hline
Thm. 7,  \cite{LKZ25}          & \(\hspace{-3mm}\geq 1\hspace{-3mm}\)          & \(\hspace{-1.5mm}d(t+1)\hspace{-1.5mm} \)             & \(O\big( N^{1/\lfloor (2d-1)/t \rfloor}\cdot \log p\big)\)                                          \\        \hline
 Thm. 1,  \cite{BGIK22}         & \(1\)               & \(4\)                   & \(O\big( 2^{2p\cdot \nu_2(N)}\cdot \log p\big) \hspace{5mm} (p\geq 3)\)                                                          \\         \hline
Thm. 8,  \cite{LKZ25}          & \(1\)               & \(8\)                   & \(O\big(2^{10\cdot \nu_2(N)} + \log p\big)\)                                                            \\         \hline
Thm. 10, \cite{LKZ25}         & \(1\)               & \(2n_r\)                 &  \(O\big( 2^{c_1(r)\cdot \nu_r(N)}\cdot \log p\big)\)     \\         \hline
$\hspace{-5mm}\text{This paper}\hspace{-5mm}$                               & \(1\)               & \(2n_r\)            & \(O \big( 2^{c_2(r) \cdot \nu_{r+1}(N)}\cdot \log p \big)\)                                  \\         \hline
    \end{tabular}
\begin{itemize}
\scriptsize 
\item[]
 $c_1(r)\approx$   the $(r+1)$th smallest prime;
\item[] $c_2(r)=\max\{p, \text{the~}(r+1)\text{th~smallest~prime}\}$
\end{itemize}
\label{tab:itdpf}
\end{table}

\vspace{-2mm}

\noindent{\bf Our contributions.}
In this paper, we 
construct a {\em perfectly} secure $(1,2n_r)$-ITDPF with output
group \(\mathbb{G}=\mathbb{Z}_{p}\), where  \(p\) can be any prime  and
$n_r$ is defined by Eq. \eqref{equ:n_value}. 
The proposed ITDPFs achieve a  subpolynomial 
key size  of $2^{O(\nu_{r+1}(N))}$, which is asymptotically smaller than 
that of \cite{LKZ25}.
Our core technical contribution
is a novel share
conversion based on the derivative-based PIR of  Ghasemi et al. \cite{GKS25}.

\section{Preliminaries}
\label{sec:Preliminaries}

\subsection{Notation}
For any integer \(n>0\), we denote \([n]=\{1,\ldots,n\}\). For any integer \(m>0\), we denote by \(\mathbb{Z}_{m}\) the ring of integers modulo \(m\). For any prime power \(q\), we denote by \(\mathbb{F}_{q}\) the finite field of order \(q\) and denote by \(\mathbb{F}_{q}^{*}\) the multiplicative group of \(\mathbb{F}_{q}\). We denote any vectors with lower case boldface letters. For any vectors \(\mathbf{u}=(u_{1},\ldots,u_{k}),\mathbf{v}=(v_{1},\ldots,v_{k})\) over a ring, we denote their inner product with \(\mathbf{u}\cdot\mathbf{v}=\sum_{i=1}^{k}u_{i}v_{i}\) and denote their pointwise product with \(\mathbf{u}\odot\mathbf{v}=(u_{1}v_{1},\ldots,u_{k}v_{k})\). We use \(\delta_{\alpha,x}\) to denote the Kronecker symbol, i.e., \(\delta_{\alpha,x}=1\) when \(x=\alpha\) and \(\delta_{\alpha,x}=0\) when \(\alpha\neq x\).

\subsection{Point functions}
Let \(N>0\) be an integer and let \(\mathbb{G}\) be an Abelian group. For any \(\alpha\in[N]\) and \(\beta\in\mathbb{G}\), the \textit{point function} \(f_{\alpha,\beta}\colon[N]\to\mathbb{G}\) is defined by \(f_{\alpha,\beta}(x)=\beta\cdot\delta_{\alpha,x}\).

%\vspace{-0.5em}
\subsection{Information-Theoretic DPF}
A \(t\)-private \(n\)-server information-theoretic DPF \cite{BGIK22} allows one to secret-share a point function \(f_{\alpha,\beta}\) among \(n\) servers such that  each server can compute an additive share of \(f_{\alpha,\beta}(x)\in\mathbb{G}\) for any given input \(x\in[N]\)
but any \(t\) servers learn absolutely no information about the function.

\begin{definition}[Distributed point function]
    A \(t\)-private \(n\)-server DPF (\((t,n)\)-DPF)  \(\Pi=(\mathsf{Gen},\{\mathsf{Eval}_{i}\}_{i=0}^{n-1})\) consists of \(n+1\) algorithms with the following syntax:

    \begin{itemize}
        \item \((k_{0},\ldots,k_{n-1})\leftarrow\mathsf{Gen}(1^{\lambda},f_{\alpha,\beta})\): Given a security parameter \(\lambda\) and a point function \(f_{\alpha,\beta}\), the (randomized) key generation algorithm returns \(n\) secret keys \(k_{0},\ldots,k_{n-1}\).
        \item \(y_{i}\leftarrow\mathsf{Eval}_{i}(k_{i},x)\): Given a secret key \(k_{i}\) and an input \(x\in[N]\), the (deterministic) evaluation algorithm \(\mathsf{Eval}_{i}\) (of server \(i\)) returns a group element \(y_{i}\in\mathbb{G}\).
    \end{itemize}

    The scheme \(\Pi\) should satisfy the following requirements:

    \begin{itemize}
        \item {\bf Correctness.} For any security parameter \(\lambda\), any point function \(f_{\alpha,\beta}\), any input \(x\in[N]\), and any secret  keys \((k_{0},\ldots,k_{n-1})\leftarrow\mathsf{Gen}(1^{\lambda},f_{\alpha,\beta})\),
              \begin{align}
                  \Pr\left[\sum_{i=0}^{n-1}\mathsf{Eval}_{i}(k_{i},x)=f_{\alpha,\beta}(x)\right]=1.
              \end{align}

        \item  {\bf Security.} Consider the following security experiment between a challenger and an adversary \(\mathcal{A}\) that controls the \(j\)-th server for all \(j\in T\) \((T\subseteq\{0,1,\ldots,n-1\},|T|\leq t)\):
              \begin{itemize}
                  \item Given the security parameter \(\lambda\), \(\mathcal{A}\) chooses two point functions \(f^{0}=f_{\alpha_{0},\beta_{0}},f^{1}=f_{\alpha_{1},\beta_{1}}: [N]\rightarrow \mathbb{G}\) and gives them to the challenger.
                  \item The challenger samples \(b\leftarrow \{0,1\}\) uniformly, generates \(n\) secret keys \((k_{0},\ldots,k_{n-1})\leftarrow\mathsf{Gen}(1^{\lambda},f^{b})\), and gives \(k_{T}=\{k_{i}:i\in T\}\) to \(\mathcal{A}\).
                  \item The adversary \(\mathcal{A}\) outputs a guess \(b^{\prime}\leftarrow\mathcal{A}(k_{T})\).
              \end{itemize}
              Let \(\operatorname{Adv}(1^{\lambda},\mathcal{A},T)=|\Pr[b=b^{\prime}]-1/2|\) be the advantage of \(\mathcal{A}\) in guessing \(b\) in the experiment. For a circuit size bound \(M=M(\lambda)\) and an advantage bound \(\epsilon=\epsilon(\lambda)\), we say that \(\Pi\) is \((M,\epsilon)\)-secure if for all subset \(T\subseteq\{0,\ldots,n-1\}\) of cardinality \(\leq t\), and all non-uniform adversaries \(\mathcal{A}\) of size \(M(\lambda)\), \(\operatorname{Adv}(1^{\lambda},\mathcal{A},T)\leq\epsilon(\lambda)\).
    \end{itemize}
\end{definition}

A \((t,n)\)-DPF  is called \textit{statistically \(\epsilon\)-secure} if it is \((M,\epsilon)\)-secure for all \(M\), and \textit{perfectly secure} if it is statistically \(0\)-secure. Both kinds of DPFs are called \textit{information-theoretic} DPFs (ITDPFs) \cite{BGIK22}. To prove that a \((t,n)\)-DPF is an ITDPF, it suffices to show that the joint distribution of any \(t\) secret keys are independent of the underlying point function, i.e., if \((k_{0},\ldots,k_{n-1})\leftarrow\mathsf{Gen}(1^{\lambda},f)\), then \(k_{T}\) is independent of \(f\) for any subset \(T\) of cardinality \(\leq t\).

\subsection{LKZ Framework for Constructing ITDPFs}
\label{subsec:DPF_Framework}
Li et al. \cite{LKZ25} has a framework that may transform certain secret sharing schemes to perfectly secure ITDPFs via share conversions \cite{BIKO12}.

\noindent{\bf Secret sharing scheme (SSS).}
An \textit{SSS}  \(\mathcal{L}=({\sf Share},{\sf Rec})\)  for \(n\) participants allows a dealer to convert a secret \(s\in\mathcal{S}\) into \(n\) shares \((\mathbf{c}_{0},\ldots,\mathbf{c}_{n-1})\leftarrow{\sf Share}(s)\) in a share space \(\mathcal{C}\), such that any authorized set 
$A\subseteq\{0,1,\ldots,n-1\}$  can reconstruct \(s\) as \(s\leftarrow{\sf Rec}(\{\mathbf{c}_{j}\}_{j\in A})\) but any unauthorized set  gives   no information about \(s\). A \((t,n)\)-\textit{threshold SSS} (TSSS) is an SSS where the authorized sets are subsets of cardinality \(> t\). An {\em additive} SSS is an \((n-1,n)\)-TSSS where the secret space is an additive group and \(s=\mathbf{c}_{0}+\cdots+\mathbf{c}_{n-1}\) is the sum of \(n\) random group elements.

\noindent{\bf Share conversion.}
Let \((\mathcal{L}_{1},\mathcal{S}_{1})=(({\sf Share}_{1},{\sf Rec}_{1}),\mathcal{S}_{1})\) and \((\mathcal{L}_{2},\mathcal{S}_{2})\) be two SSSs. Let \(R\subseteq\mathcal{S}_{1}\times\mathcal{S}_{2}\) be a binary relation such that, for every \(s_{1}\in\mathcal{S}_{1}\) there exists an \(s_{2}\in\mathcal{S}_{2}\) satisfying \((s_{1},s_{2})\in R\). The scheme \(\mathcal{L}_{1}\) is locally convertible to \(\mathcal{L}_{2}\) w.r.t. \(R\) if there exist \(n\)  {\em share conversion} functions \(g_0,\ldots,g_{n-1}:\mathcal{C}_{1}\to\mathcal{C}_{2}\) such that: for any \(s_{1}\in\mathcal{S}_{1}\) and any \((\mathbf{c}_{0},\ldots,\mathbf{c}_{n-1})\leftarrow{\sf Share}_{1}(s_{1})\), \((g_{0}(\mathbf{c}_{0}),\ldots,g_{n-1}(\mathbf{c}_{n-1}))\) is a valid sharing of an \(s_{2}\in\mathcal{S}_{2}\) that satisfies  \((s_{1},s_{2})\in R\). 

\noindent{\bf The framework.}
Let \((\mathcal{L}_{1},\mathcal{S}_{1})=(({\sf Share}_{1},{\sf Rec}_{1}),[N])\) be a \((t,n)\)-TSSS with share space \(\mathcal{C}_{1}\) and let \((\mathcal{L}_{2},\mathcal{S}_{2})\) be an additive SSS with share space 
  \(\mathcal{C}_{2}=\mathcal{S}_{2}\). Suppose that \((\mathcal{L}_{1},\mathcal{S}_{1})\) is locally convertible to \((\mathcal{L}_{2},\mathcal{S}_{2})\) w.r.t. a binary relation \(R\subseteq\mathcal{S}_{1}\times\mathcal{S}_{2}\). To obtain an \(n(t+1)\)-server ITDPF with domain \([N]\) and output group \(\mathbb{G}\), the framework   \cite{LKZ25} requires:

\begin{itemize}
    \item[(a)] There is a function \(\mathsf{Conv}:\{0,1,\ldots,n-1\}\times\mathcal{S}_{1}\times\mathcal{C}_{1}\to\mathcal{C}_{2}\) such that for any \(x\in\mathcal{S}_{1}\), \(\{g_{\ell}^{x}=\mathsf{Conv}(\ell,x,\cdot)\}_{\ell=0}^{n-1}:\mathcal{C}_{1}\to\mathcal{C}_{2}\) are \(n\)  share conversion functions for \(R\).
    \item[(b)] There is a commutative ring \(\mathbb{R}\) with identity \(1\) such that \(\mathbb{G}\) is a subgroup of the additive group of \(\mathbb{R}\), \(1\in\mathbb{G}\), and there is a surjective homomorphism \(\phi:\mathbb{R}\to\mathbb{G}\).
    \item[(c)] There exist an \(\mathbb{R}\)-module \(\mathbb{H}\), a function \(\psi:\mathcal{S}_{1}\rightarrow\mathbb{H}\), and a bilinear function \(\Phi:\mathbb{H}\times\mathcal{C}_{2}\rightarrow\mathbb{R}\) such that: for any \(\alpha,x\in\mathcal{S}_{1}\), \((\mathbf{c}_{0},\ldots,\mathbf{c}_{n-1})\leftarrow{\sf Share}_{1}(\alpha)\), and
          \begin{align}
              \label{equ:rho_def}
              \rho(\alpha,x)=\Phi\left(\psi(\alpha),\sum_{\ell=0}^{n-1}\mathsf{Conv}(\ell,x,\mathbf{c}_{\ell})\right),
          \end{align}
          there exists a ring element \(\sigma\in\mathbb{R}\) that satisfies
          \begin{align}
              \label{equ:Requirement_c}
              \phi(\rho(\alpha,x)\cdot\sigma)=\delta_{\alpha,x}.
          \end{align}
\end{itemize}

Given any point function \(f_{\alpha,\beta}:[N]\rightarrow \mathbb{G}\), the framework \cite{LKZ25} generates \(k=n(t+1)\) keys by first computing \(n\) shares \((\mathbf{c}_{0},\ldots,\mathbf{c}_{n-1})\leftarrow{\sf Share}_{1}(\alpha)\) of \(\alpha\) under \(\mathcal{L}_{1}\), then additively sharing \((\sigma\cdot\beta)\circ\psi(\alpha)\) as the sum of \(t+1\) random  module elements \(\omega_{0},\omega_1,\ldots,\omega_{t}\in\mathbb{H}\), i.e.,
\begin{align}
    \omega_{0}+\omega_{1}+\cdots+\omega_{t}=(\sigma\cdot\beta)\circ\psi(\alpha),
\end{align}
where \(\circ\) is the scalar multiplication between ring elements and module elements, and finally setting the \(i\)-th key as
\begin{align}
    k_{i}=(\omega_{j},\mathbf{c}_{\ell})
\end{align}
for all \(i=jn+\ell\), where \(0\leq j\leq t\) and \(0\leq\ell\leq n-1\).
Given the key \(k_{i}=(\omega_{j},\mathbf{c}_{\ell})\) and any input \(x\in[N]\), the 
 \(i\)-th server's  evaluation algorithm simply outputs
\begin{align}
    y_{i}=\phi(\Phi(\omega_{j},\mathsf{Conv}(\ell,x,\mathbf{c}_{\ell}))).
\end{align}

%\vspace{-1em}
\subsection{GKS Derivative-based PIR}
\label{sec:PIR_Scheme}
A (1-private) \(k\)-server PIR scheme allows one to retrieve any element \(\lambda_{\alpha}\) of a database \(\bm \lambda=(\lambda_1,\ldots,\lambda_N)\) from \(k\) servers, where each server has a copy of $\bm \lambda$, without revealing   \(\alpha\in[N]\) to each individual server. Recently, Ghasemi et
al.  \cite{GKS25} presented the state-of-the-art 1-private PIR by improving 
the schemes of \cite{CFLWZ13,DG15,E09}, which are all based on \(S\)-matching families and \(S\)-decoding polynomials.

\begin{definition}[\(S\)-matching  family]
    For any integers \(m,h,N>0\) and any set \(S\subseteq\mathbb{Z}_{m}\) with \(0\in S\), an \(S\)-matching family of size \(N\) in \(\mathbb{Z}_{m}^{h}\) is a pair \((\mathcal{U}=\{\mathbf{u}_{i}\}_{i=1}^{N},\mathcal{V}=\{\mathbf{v}_{i}\}_{i=1}^{N})\) of subsets of 
    $\mathbb{Z}_{m}^{h}$ that satisfies the following properties:
    \begin{itemize}
        \item \(\mathbf{u}_{i}\cdot\mathbf{v}_{i}=0\) for all \(i\in[N]\); and
        \item \(\mathbf{u}_{i}\cdot\mathbf{v}_{j}\in S\setminus\{0\}\) for all \(i\neq j\).
    \end{itemize}
\end{definition}

\begin{definition}[\(S\)-decoding polynomial]
    For any integer \(m>0\), any set \(S\subseteq\mathbb{Z}_{m}\) with \(0\in S\), any prime \(q\) such that \(m|(q-1)\) and a  primitive \(m\)-th root of unity  \(\gamma\in \mathbb{F}_{q}\), an \(S\)-decoding polynomial is a polynomial \(P(X)\in\mathbb{F}_{q}[X]\) such that
    \begin{itemize}
        \item \(P(1)=1\); and
        \item \(P(\gamma^{s})=0\) for all \(s\in S\setminus\{0\}\).
    \end{itemize}
\end{definition}

For any integer \(m=p_{1}\cdots p_{r}\) that is a product of   \(r\geq 2\) distinct primes, Grolmusz's set systems \cite{G00} gave an \(S_{m}\)-matching family of size 
\begin{align} 
  N=\exp(O((\log h)^r/(\log\log h)^{r-1}))
\end{align}
 in \(\mathbb{Z}_{m}^{h}\), where     $S_m$ is the 
    {\em canonical set}  of \(m\) and defined as
\begin{align} S_{m}=\{s\in\mathbb{Z}_{m}:s\bmod p_{i}\in\{0,1\},\forall i\in[r]\}.
\end{align} 
With these families and a trivial construction  of \(S\)-decoding polynomials, Efremenko \cite{E09} constructed a $(1,2^r)$-PIR  of subpolynomial communication complexity
 $\exp(O(\nu_r(N)))$. In  \cite{E09}, the database \(\bm \lambda\) is  encoded as
an $h$-variate  polynomial  
\begin{align}
    \label{equ:data-encoding-poly}
    G_{\bm \lambda}(\mathbf{z})=\sum_{j=1}^{N}\lambda_{j}\mathbf{z}^{\mathbf{u}_{j}}:H_{m}^{h}\rightarrow\mathbb{F}_{q};
\end{align}
over $H_m\subseteq \mathbb{F}_q^*$, the order-$m$ subgroup of 
$\mathbb{F}_q^*$ for a prime power $q$ that satisfies $m|(q-1)$, 
each server is queried with a specific point \(C(b_\ell)\) on a random multiplicative  line \(C(Z)=\mathbf{w}\odot Z^{\mathbf{v}_{\alpha}}\) in $H_m^h$ and responds with an evaluation of the function
\begin{align} 
    g(Z)=G_{\bm \lambda}(\mathbf{w}\odot Z^{\mathbf{v}_{\alpha}})
\end{align}
at {\(b_\ell\)}, and finally   \(\lambda_{\alpha}=g(0)\) is recovered by interpolating
 \(g(Z)\). 
 
 While subsequent works \cite{CFLWZ13,DG15} significantly reduced the number of servers required by \cite{E09}, the recent work \cite{GKS25}   further reduced this  number   by asking each server to provide not only evaluations of \(G_{\bm \lambda}\) but also its Hasse derivatives,
    which allow \cite{GKS25}  to use an \(S_{M}\)-matching family to construct the polynomial in (\ref{equ:data-encoding-poly}) but an \(S_{M}\)-decoding polynomial that contains no more terms
    than the sparsest  \(S_{m}\)-decoding polynomial,\
where \(M=mp\) for a prime \(p\) such that \(\gcd(m,p)=1\). A main technical contribution of \cite{GKS25} is the   study of the \(0\)-interpolation property with multiplicity.

\begin{definition}[\(0\)-interpolation property with multiplicity \(e\)]
    Let \(S\subseteq\mathbb{N}\) with \(0\in S\). A set \(B\subseteq\mathbb{F}_{q}\) has the \(0\)-interpolation property with multiplicity \(e\) for \(S\) if there exists a linear map \(E:(\mathbb{F}_{q}^{e})^{B}\rightarrow\mathbb{F}_{q}\) such that \(E(R^{(<e)}|_{B})=R(0)\) for any polynomial \(R(Z)\) of the form \(R(Z)=\sum_{s\in S}c_{s}Z^{s}\), where \(R^{(<e)}|_{B}\) denotes the vector of  the evaluations of
    the \(i\)-th Hasse derivatives of \(R\) at \(B\) for all \(i<e\).
\end{definition}

Let \(\mathbb{F}\) be a finite field of characteristic \(p\) such that \(m| (|\mathbb{F}|-1)\)
and let \(H_m\subseteq \mathbb{F}\) be the order-$m$ subgroup of \(\mathbb{F}^*\).
Ghasemi et al.  \cite{GKS25} showed that {\em if  $0\in S_M$ and \(S_M \subseteq \varphi(S_m \times \{0,1,\ldots,e-1\})\) for the
Chinese Remainder Isomorphism  \(\varphi: \mathbb{Z}_m \times \mathbb{Z}_p \rightarrow \mathbb{Z}_M\), then
any \(0\)-interpolating set \(B\subseteq H_m\) with multiplicity 1 for \(S_{m}\) must be
a \(0\)-interpolating set with multiplicity $e$ for \(S_{M}\).}
Based on this property, Ghasemi et al.  \cite{GKS25} obtained a 1-private \(n_r\)-server PIR of communication complexity  \(2^{O(\nu_{r+1}(N))}\) for any \(r\geq 1\), where
$n_r$ is defined by Eq. \eqref{equ:n_value}.

\section{Our DPF Construction}
\label{sec:construction}

This section presents our new $(1,2n_r)$-ITDPF  with shorter keys of size 
$2^{O(\nu_{r+1}(N))}$. The proposed construction  is based on the LKZ framework \cite{LKZ25}
and consists of two steps. The first of these steps  extracts the required  share conversions out of the GKS  
derivative-based PIR \cite{GKS25} and consists of our main technical contribution; 
and the second step  converts the resulting share conversions to the proposed ITDPF, using
the idea of \cite{LKZ25}.

\subsection{From GKS Derivative-based PIR to Share Conversion}
\label{sec:params}

Let \(m=p_1\cdots p_r\) be the product of \(r \geq 2\) distinct primes and let \(M=mp\)
for a prime \(p\) such that \(\gcd(m,p)=1\).
Let \(\mathbb{F}\) be a finite field of  size $p^{\tau}$ such that \(m \mid(|\mathbb{F}|-1)\).
Let  \(H_m\) be the order-$m$ subgroup of  \(\mathbb{F}^*\). 
Let \(S_m \subseteq \mathbb{Z}_m\)
and \(S_M\subseteq \mathbb{Z}_M\) be the canonical sets of \(m\) and \(M\), respectively.
By Theorem 1.4 of \cite{G00}, there is an
\(S_M\)-matching  family \((\mathcal{U}=\{\mathbf{u}_{i}\}_{i=1}^{N},\mathcal{V}=\{\mathbf{v}_{i}\}_{i=1}^{N}) \) in $\mathbb{Z}_M^h$   for
properly chosen $h$.
By Theorem 4.1 of \cite{CFLWZ13}, there is an \(S_m\)-decoding polynomial over \(\mathbb{F}\) with  \( n_r\) terms
for the \(n_r\) in (\ref{equ:n_value}), i.e.,
there  is a 0-interpolating set 
\begin{align} 
\label{eqn:B}
B=\{b_0,b_1,\ldots,b_{n_r-1}\} \subseteq H_m
\end{align}
 with multiplicity 1 for \(S_m\).
 Note that $S_M\subseteq \varphi(S_m\times \{0,1\})$,  \(B\) must be a 0-interpolating  set with  multiplicity 2 for  \(S_M\).

\vspace{2mm}
\noindent{\bf Choosing  SSSs.}
Let \(f_{\alpha,\beta}:[N]\rightarrow \mathbb{Z}_p\) be a point function with  output group \( \mathbb{Z}_p\).  The proposed share conversion chooses an SSS   \((\mathcal{L}_{1},\mathcal{S}_{1})=(({\sf Share}_{1},{\sf Rec}_{1}),{\cal S}_1)\) with share space ${\cal C}_1$ and an SSS \((\mathcal{L}_2,\mathcal{S}_2)\) with share space ${\cal C}_2$ such that
\begin{align}
    \label{equ:SSS_def}
    \mathcal{S}_1=[N],\ \ \ \mathcal{C}_1 = \mathbb{F}^{h+1},\ \ \ \mathcal{S}_2=\mathcal{C}_2=\mathbb{F}^{h+1}
\end{align}
For  \(\alpha \in \mathcal{S}_1\), \({\sf Share}_1(\alpha)\) generates \(n\) shares \(\mathbf{c}_0,\mathbf{c}_1,\ldots,\mathbf{c}_{n_r-1} \in \mathcal{C}_1\) by mapping \(\alpha\) to the vector \(\mathbf{v}_{\alpha} \in \mathcal{V}\), randomly choosing a vector \(\mathbf{w} \in H_m^h\) and finally setting
\begin{align} \label{eqn:cl}
    \mathbf{c}_{\ell} = (\mathbf{w} \odot b_{\ell}^{\mathbf{v}_{\alpha}},b_{\ell}),\ \ \ell=0,1,\cdots,n_r-1.
\end{align}
The SSS \({\cal L}_1\) is 1-private because the first \(h\) entries of any share \({\bf c}_\ell\)
are truly random and independent of \(\alpha\). However, given any two shares
\(\mathbf{c}_{\ell_1}\) and \(\mathbf{c}_{\ell_2}\), \( (b_{\ell_1}/b_{\ell_2})^{\mathbf{v}_{\alpha}}\) can be easily extracted  by dividing the first \(h\) entries of both shares and then give full information
about  \(\mathbf{v}_{\alpha}\) (and so $\alpha$), because 
both \(b_{\ell_1}\) and \(b_{\ell_2}\) are public.
The SSS \(\mathcal{L}_2\) is simply an additive SSS over \(\mathcal{S}_2\).

\vspace{2mm}
\noindent
{\bf Constructing $\sf Conv$.}
For every \(\ell \in \{0,1,\ldots,n_r-1\}\), we need to construct a function that enables the \(\ell\)th server  to transform its share \(\mathbf{c_{\ell}}\) into a component that eventually 
contributes to the  reconstruction of the point function \(f_{\alpha,\beta}\).
Given the share \(\mathbf{c_{\ell}}\) and any input \(x \in \mathcal{S}_1\), the local share conversion function
\({\sf Conv}(\ell,x,\mathbf{c_{\ell}})\) is designed in four steps as follows.

\vspace{2mm}
\subsubsection{Defining the core polynomial \({D}_x(Z)\)}
We begin by associating a specific univariate polynomial
\begin{align}
    \label{equ:DxZ}
    {D}_x(Z) = \sum_{j=1}^{N}f_{j,1}(x)\cdot \mathbf{w}^{\mathbf{u}_j}Z^{\mathbf{v}_{\alpha}\cdot\mathbf{u}_j }
\end{align}
over the finite field \(\mathbb{F}\)
with the  input \(x\).
In fact, if we denote ${\bm \lambda}_x=(f_{1,1}(x),\ldots,f_{N,1}(x))$, then the core polynomial
$D_x(Z)$ can be obtained  by restricting the function 
\begin{align} \label{eqn:Flxz}
F_x({\bf z}):=G_{\bm \lambda_x}(\mathbf{z})=\sum_{j=1}^{N}f_{j,1}(x)\mathbf{z}^{\mathbf{u}_j}
\end{align}
 defined by  Eq. 
(\ref{equ:data-encoding-poly}) 
 to the multiplicative line 
 \begin{align}
 C(Z) = \mathbf{w} \odot Z^{\mathbf{v}_{\alpha}}.
 \end{align}
Note that the polynomial
\(\tilde{D}_{x}(Z) = D_x(Z) \bmod (Z^M-1)\) admits a sparse representation of the form
\begin{align}
    \label{equ:D_1}
    \tilde{D}_{x}(Z) = f_{\alpha,1}(x)\cdot \mathbf{w}^{\mathbf{u}_{\alpha}} + \sum_{s \in S_M \backslash \{0\}} k_s Z^s,
\end{align}
due to the the properties of \(S_M\)-matching family, where
\begin{align}
k_s=\sum_{j\in[N]: \mathbf{v}_{\alpha}\cdot\mathbf{u}_j=s} f_{j,1}(x)\cdot {\bf w}^{{\bf u}_j}
\end{align}
for all \(s\in S_M\setminus\{0\}\).
The reduced form \(\tilde{D}_x(Z)\) allows one to compute \(f_{\alpha,1}(x)\) as
\begin{align}\label{eqn:fa1}
f_{\alpha,1}(x)=\mathbf{w}^{-\mathbf{u}_{\alpha}}\cdot \tilde{D}_{x}(0)
\end{align}
and thus reduce the problem of computing of
\(f_{\alpha,1}(x)\) to that of recovering \(\tilde{D}_{x}(0)\).

\vspace{2mm}
\subsubsection{Applying 0-interpolation property with multiplicity}
Recall that the set \(B\) in Eq. \eqref{eqn:B} is a 0-interpolating set  with multiplicity 2 for
 \(S_M\)
and  guarantees a linear map 
\begin{align}\label{eqn:E}
E:\ (\mathbb{F}^2)^B \rightarrow \mathbb{F}
\end{align} such that  the constant term \(\tilde{D}_{x}(0)\) can be recovered from the   evaluations and first-order derivatives of \(\tilde{D}_{x}(Z)\) at all points in \(B\), i.e. there exist constants \(\{a_{\ell,k}: 0\leq \ell<n_r, k=0,1\}\) such that
\begin{align} \label{eqn:Dx0}
    \tilde{D}_{x}(0) = \sum_{\ell=0}^{n_r-1} \sum_{k=0}^{1} a_{\ell,k}\cdot  \tilde{D}_{x}^{{(k)}}(b_{\ell}).
\end{align}
As \(\mathbb{F}\) is a field of characteristic \(p\),  \((Z-b_{\ell})^p|(Z^M-1)\) for all \(\ell\in \{0,\ldots,n_r-1\}\).
It  follows that \(\tilde{D}_{x}\) and \(D_x\) agree at each \(b_{\ell}\) not only in their evaluations,   but also in their first \((p-1)\) derivatives. In particular,
\begin{align}\label{eqn:Dx1}
    \tilde{D}_{x}(b_{\ell}) = {D}_x(b_{\ell}), \ \tilde{D}_{x}^{(1)}(b_{\ell}) = D_x^{(1)}(b_{\ell}),~~ 0\leq \ell<n_r.
\end{align}
Based on (\ref{eqn:Dx0}) and (\ref{eqn:Dx1}), the problem of recovering \(\tilde{D}_{x}(0)\) can be reduced to that of computing the evaluations and first-order derivatives of   \(D_x(Z)\) at all points in \(B\).

\vspace{2mm}
\subsubsection{Computing   derivatives via chain rule}
While \(D_x(b_\ell)\) can be easily computed   by  each server,  \(D_x^{(1)}(b_{\ell})\) must be recovered from the Hasse derivatives of
a multivariate polynomial.
Recall that $D_x(Z)=F_x(C(Z))$, we must have
\begin{align}
    \label{D_Hasse}
    D_x^{(1)}(Z) = \langle \nabla F_x(C(b_{\ell})),C^{(1)}(b_{\ell}) \rangle
\end{align}
where for all $b_\ell\in B$,
\begin{align}
\nabla F_x(C(b_\ell)) & = (\partial_1F_x({\bf z}), \ldots,  \partial_hF_x({\bf z}) )|_{{\bf z}=C(b_\ell)} \label{eqn:Fxcbl} \\
C^{(1)}(b_{\ell})&=b_{\ell}^{-1}  C(b_{\ell})\odot \mathbf{v}_{\alpha}.\label{eqn:C1bl}
\end{align}
Thus, for all  \(b_\ell\in B\), to compute
the derivative  $D_x^{(1)}(b_{\ell})$, it suffices for 
 the client to learn $\nabla F_x(C(b_\ell)$ from a server.

\vspace{2mm}
\subsubsection{Embedding the linear map coefficient}
Recall that the linear map  $E$ in Eq. \eqref{eqn:E}    ensures  the recovery of   \(\tilde{D}_{x}(0)\)
from the values in Eq. \eqref{eqn:Dx1}, with coefficients 
 \(\{a_{\ell,k}: 0\leq \ell<n_r, k=0,1\}\) from Eq. \eqref{eqn:Dx0}.
Given that ${\bf c}_\ell=(b_\ell^{-1}C(b_\ell), b_\ell)$,  our design of   \({\sf Conv}\) is 
  finally completed by 
embedding the recovery coefficients into Eq. \eqref{eqn:Fxcbl} (which are required by 
Eq. (\ref{D_Hasse}))
and given by
\begin{align}
    \label{equ:Share_Conversion}
    \hspace{-7mm}   {\sf Conv}(\ell,x,\mathbf{c_{\ell}})\hspace{-0.5mm}=\hspace{-0.5mm}
    \Big(a_{\ell,0} D_x(b_{\ell}), \hspace{-1mm}\ a_{\ell,1}\frac{C(b_{\ell})}{b_{\ell}} \odot \nabla F_x(C(b_{\ell})) \Big).
    \hspace{-5mm}
\end{align}
Given  \eqref{equ:Share_Conversion}, the binary relation $R$ in our DPF  
can be described as follows.
For  $s_1=\alpha\in {\cal S}_1$,   \(\mathbf{w}\in H_m^h\),   \(x \in [N]\),
and   the ${\bf c}_\ell$'s in Eq. \eqref{eqn:cl}, 
the secret $s_2\in {\cal S}_2$ such that $(s_1,s_2)\in R$ is 
the sum of the $n_r$ converted shares in Eq. \eqref{equ:Share_Conversion}:
\begin{align}
    \label{equ:SSS2_Rec}
    \mathbf{s}_2(\mathbf{w}, \alpha, x) = \sum_{\ell=0}^{n_r-1} {\sf Conv}(\ell, x, \mathbf{c}_\ell).
\end{align}

\subsection{From Share Conversion to DPF}
\label{sec:dpf-algos}

So far the $(1,n_r)$-TSSS \((\mathcal{L}_1,\mathcal{S}_1)\) (Eq. \eqref{equ:SSS_def}, \eqref{eqn:cl}) and 
 the additive SSS \((\mathcal{L}_2,\mathcal{S}_2)\) (Eq. \eqref{equ:SSS_def}, \eqref{equ:SSS2_Rec}) we have chosen  in 
Section \ref{sec:params} and the new share  conversion function $\sf Conv$ (Eq. \eqref{equ:Share_Conversion})  we have constructed in Section  \ref{sec:params} have been 
confirmed to satisfy the property (a) required by the LKZ framework (see  Section \ref{subsec:DPF_Framework}). 
To complete the DPF construction, we need to choose  a ring \(\mathbb{R}\),  an \(\mathbb{R}\)-modulo \(\mathbb{H}\) and three functions \(\phi,\psi,\Phi\) such that (b) and (c) are also satisfied.

\vspace{2mm}
\noindent{\bf Choosing $(\mathbb{R},\phi)$.}
Note that   \(\mathbb{F}=\mathbb{F}_{p^\tau}\) is a field of characteristic \(p\).
There exists a monic  irreducible polynomial $\zeta(X)\in \mathbb{Z}_p[X]$ of degree \(\tau\) such that 
 \(\mathbb{F} \cong \mathbb{Z}_p[X]/(\zeta(X))\).
Then every field element  $y\in \mathbb{F}$ has the form 
 $y =y_0+y_1X+\cdots +y_{\tau-1}X^{\tau-1}$.  
We choose  $\mathbb{R}=\mathbb{F}$ and choose  
  $\phi: \mathbb{R}\rightarrow \mathbb{G}$ such that 
\begin{align}
    \phi(y)=y_0,  \ \ \ \forall y\in \mathbb{F}.
\end{align}
It is trivial to verify that  $(\mathbb{R},\phi)$ is  a surjective homomorphism
and the property (b) required by   LKZ is satisfied.

\vspace{2mm}
\noindent{\bf Choosing $(\mathbb{H},\psi, \Phi)$.}
To meet the requirement (c), we choose the \(\mathbb{R}\)-modulo \(\mathbb{H}=\mathbb{F}^{h+1}\), choose   \(\psi:\mathcal{S}_1 \rightarrow \mathbb{H}\) such that
\begin{align}
    \label{equ:psi}
    \psi(\alpha) = (1,\mathbf{v}_{\alpha}),\ \ \ \forall\ \alpha \in \mathcal{S}_1,
\end{align}
and choose the bilinear function \(\Phi:\mathbb{H} \times \mathcal{C}_2 \rightarrow \mathbb{R}\) such that 
\begin{align}
    \label{equ:Phi}
    \Phi(h,c) = \langle h,c \rangle,\ \ \forall\ h  \in \mathbb{H}, c\in {\cal C}_2.
\end{align}
Then for  any \(\alpha, x \in \mathcal{S}_1\), \((\mathbf{c}_0,\mathbf{c}_1,\ldots,\mathbf{c}_{n_r-1}) \leftarrow {\sf Share}_1(\alpha)\), the equations (\ref{equ:SSS2_Rec}), (\ref{equ:psi}) and (\ref{equ:Phi})
jointly imply that the $\rho(\alpha,x)$ in Eq. \eqref{equ:rho_def} has the following form
\begin{align}
    \rho(\alpha,x) = \langle (1,\mathbf{v}_{\alpha}),s_2(\mathbf{w},\alpha,x) \rangle
\end{align}
By Eq.
\eqref{eqn:Dx0}, \eqref{eqn:Dx1}, \eqref{D_Hasse}, \eqref{eqn:Fxcbl}, \eqref{eqn:C1bl},
\eqref{equ:Share_Conversion}, and \eqref{equ:SSS2_Rec}, it is easy to see that $\rho(\alpha,x)=  \tilde{D}_{x}(0)$. 
Considering  \eqref{eqn:fa1},  we can easily satisfy 
 (\ref{equ:Requirement_c}) by further choosing
\begin{align}
\sigma=\mathbf{w}^{-\mathbf{u}_{\alpha}}
\end{align}
and thus eventually meet the requirement (c). 

\vspace{2mm}
\noindent{\bf The $(1,2n_r)$-ITDPF.}
Finally, by applying the transformation from Section \ref{subsec:DPF_Framework} we 
get a  perfectly secure \((1,2n_r)\)-ITDPF (Fig. 1) with output group \(\mathbb{Z}_p\), 
where  \(n_r\) is defined by  (\ref{equ:n_value}).
\begin{figure}[H]
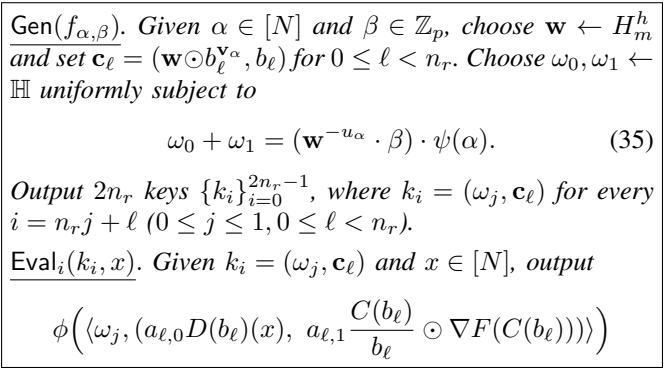

    \centering
    \em
    \begin{boxedminipage}{8.8cm}
       \(\underline{{\sf Gen}(f_{\alpha,\beta})}\). Given $\alpha\in[N]$ and $ \beta\in \mathbb{Z}_p$, choose
       ${\bf w}\leftarrow H_m^h$ and set  \(\mathbf{c}_{\ell} = (\mathbf{w} \odot b_{\ell}^{\mathbf{v}_{\alpha}},b_{\ell})\) for \(0\leq \ell<n_r \).
        Choose \(\omega_0,\omega_1 \leftarrow \mathbb{H}\) uniformly subject to
        \begin{align}
            \label{equ:share_h}
            \omega_0+\omega_1 = (\mathbf{w}^{-u_{\alpha}} \cdot \beta)  \cdot \psi(\alpha).
        \end{align}
        Output \(2n_r\) keys \(\{k_i\}_{i=0}^{2n_r-1}\), where \(k_i=(\omega_j,\mathbf{c}_{\ell})\) for every \(i = n_rj + \ell\) (\(0 \leq j \leq 1, 0 \leq \ell<n_r\)).

\vspace{1mm}
      \(\underline{{\sf Eval}_i(k_i, x)}.\) Given  \(k_i=(\omega_j,\mathbf{c}_\ell)\) and $x\in [N]$,  output
        \[
            \phi \Big(\langle \omega_j,(a_{\ell,0} D(b_{\ell})(x),\ a_{\ell,1} \frac{C(b_{\ell})}{b_{\ell}} \odot \nabla F(C(b_{\ell})) ) \rangle \Big)
        \]
    \end{boxedminipage}
    \caption{Our $(1,2n_r)$-ITDPF}
    \label{fig:ourdpf}
\end{figure}

\subsection{Analysis}
\label{sec:correctness}

\noindent
{\bf Correctness.}
Within the LKZ   Framework it suffices to verify (a), (b) and (c) are all satisfied 
by the proposed constructions. While (a) and (b) have be confirmed, we focus 
on Eq.  (\ref{equ:Requirement_c}), i.e.,
$
    \phi(\rho(\alpha,x) \cdot \sigma) = \delta_{\alpha,x},
$
which is true because
\begin{align*}
     & \ \phi(\rho(\alpha,x) \cdot \sigma) = \phi \Big(\mathbf{w}^{-\mathbf{u}_{\alpha}} \big\langle (1,\mathbf{v}_{\alpha}),\sum_{\ell=0}^{n_r-1}
    {\sf Conv}(\ell,x,\mathbf{c}_{\ell}) \big\rangle \Big)                                                                                                                                                      \\
     & = \phi \Big(\frac{\sum_{\ell = 0}^{n_r-1} a_{\ell,0} D_x(b_{\ell}) +\mathbf{v}_{\alpha} (a_{\ell,1} \frac{C(b_{\ell})}{b_{\ell}} \odot \nabla F_x(C(b_{\ell})))}{\mathbf{w}^{\mathbf{u}_{\alpha}}} \Big) \\
     & = \phi \Big(\mathbf{w}^{-\mathbf{u}_{\alpha}}\sum_{\ell=0}^{n_r-1} \sum_{k=0}^{1} a_{\ell,k} D_x^{(k)}(b_{\ell}) \Big) = \delta_{\alpha,x}.
\end{align*}

\vspace{2mm}
\noindent
{\bf Security.}
For any   $\alpha \in [N],\beta\in \mathbb{Z}_p$ and $i=jn_r+\ell$, the
$i$-th DPF secret key 
  $k_i = (\omega_j, \mathbf{c}_\ell)$
  for the point function $f_{\alpha,\beta}$
  is completely determined by Eq. 
  \eqref{eqn:cl} (for ${\bf c}_\ell$) and Eq. \eqref{equ:share_h} (for $\omega_j$).  
Since $b_\ell$ is a public parameter, the only component of     $\mathbf{c}_\ell$ that may be relevant to the
point function $f_{\alpha,\beta}$ is $\mathbf{w} \odot b_{\ell}^{\mathbf{v}_{\alpha}}$.
This component is generated by a 
 $(1, n_r)$-TSSS and  uniformly distributed over the 
 set $H_m^h$, as $\bf w$ is uniformly chosen from $H_m^h$. 
 On the other hand, 
$\omega_j$ is an additive share of $\frac{\beta}{\mathbf{w}^{\mathbf{u}_\alpha}} \cdot \psi(\alpha)$
and uniformly distributed over the set $\mathbb{H}$.
Considering that the sharing processes of  $\mathbf{c}_\ell$ 
and $\omega_j$ are independent of each other, the key 
$k_i$ is uniformly distributed over the set $H_m^h\times \mathbb{H}$ and thus
gives no information about $f_{\alpha,\beta}$, i.e.,
the proposed DPF is perfectly secure.

\vspace{2mm}
\noindent{\bf Key size.}
Referring to Fig. \ref{fig:ourdpf}, every secret key \(k_i=(\omega_j,{\bf c}_\ell)\) consists of
an element of $\mathbb{H} = \mathbb{F}^{h+1}$ 
and an element of   $H_m^{h+1}$ and thus has size
$O(h\log p)$, where $h$ is an integer such that there is an 
$S_M$ matching family of size $N$.  Since $M=mp$ is the product of $r+1$ distinct primes, 
by \cite{G00}, we may choose  $h=2^{c_2(r)\cdot \nu_{r+1}(N)}$, where $c_2(r)$
is the larger one between $p$ and the $(r+1)$th smallest prime. 
 Hence, the key size of the proposed DPF is 
$O(2^{c_2(r)\cdot \nu_{r+1}(N)} \cdot \log p)$.

\subsection{Comparisons}
\label{sec:detailed_comparison}

Table \ref{tab:dpf-comparison} give detailed comparisons between the   $(1,2n_r)$-ITDPF
proposed by this work and the perfectly secure   $(t,n)$-ITDPFs for the
same output group $\mathbb{Z}_p$ prior to this work. 
 Compared with the constructions in 
 Thm. 7 of  \cite{LKZ25}, 
Thm. 8 of \cite{LKZ25}, and Thm. 10 of \cite{LKZ25}, our ITDPFs
achieve a key size that is asymptotically smaller as a function of $N$,
the domain size of the point function. 
Compared with  the construction in Thm. 1 of  \cite{BGIK22}, our key size is
not worse.  
 
 \section{Conclusions}

In this  paper we  proposed   a novel share conversion   based on the state-of-the-art
1-private $n_r$-server private information retrieval (PIR) of Ghasemi, Kopparty, and Sudan (STOC 2025)
and   gave    a perfectly secure $(1,2n_r)$-ITDPF  with output group \(\mathbb{G}=\mathbb{Z}_p\) by using
the LKZ framework, where
      $p$ can be any prime.   
    Compared with the best existing perfectly secure ITDPFs for the same output group,
the proposed ITDPF  is  more efficient with asymptotically shorter secret keys.

As a restriction, the proposed construction is only applicable to 
a {\em  prime-order} output group $\mathbb{Z}_p$. 
We can partially remove this restriction on output groups by 
extending the proposed construction to 
any output group  $\mathbb{G}$ that is isomorphic to 
\( \mathbb{Z}_{p_1} \times \mathbb{Z}_{p_2} \times \cdots \times \mathbb{Z}_{p_k}\), the direct product of
$k$ prime-order output groups, 
in a straightforward way. 
It is left for future work to extend the proposed 
construction to arbitrary output groups.

As another restriction,   the proposed construction gives 
{\em 1-private} DPFs only,   because the  PIR underlying  our construction    is 1-private.
Note that Barkol et al. \cite{BIW10}  has a  generic method of  transforming
any $(1,k)$-PIR to  a $(t,k^t)$-PIR. 
By applying this method to the PIR of GKS, the privacy of our ITDPF may be boosted, at the price of an 
 exponential    increase in the number of servers to \((2n_r)^t\), which however is inefficient. 
It is also an interesting future work to extend the proposed 
construction or even the LKZ framework  to support 
stronger privacy requirements.

\end{document}